 \documentclass[pmlr,twocolumn,10pt]{jmlr} 
\usepackage{rotating}

\usepackage{booktabs}
\usepackage[load-configurations=version-1]{siunitx} 

\theorembodyfont{\upshape}
\theoremheaderfont{\scshape}
\theorempostheader{:}
\theoremsep{\newline}

\jmlrvolume{ML4H Extended Abstract Arxiv Index}
\jmlryear{2020}
\jmlrsubmitted{2020}
\jmlrpublished{}
\jmlrworkshop{Machine Learning for Health (ML4H) at NeurIPS 2020} 

\title[DL Derived Image Score for Increasing Phase 3 POS]{Deep Learning Derived Histopathology Image Score for Increasing Phase 3 Clinical Trial Probability of Success\titlebreak }

\author{%
\Name{Qi Tang} \Email{qi.tang@sanofi.com}\\
\addr Digital and Data Sciences \\
    Sanofi. Bridgewater, NJ, USA
\AND
\Name{Vardaan Kishore Kumar} \Email{vkishore@stevens.edu}\\
\addr Department of Computer Science\\
   Steven Institute of Technology. Hoboken, NJ, USA
}

\begin{document}

\maketitle
\begin{abstract}
Failures in Phase 3 clinical trials contribute to expensive cost of drug development in oncology. To drastically reduce such cost, responders to an oncology treatment need to be identified early on in the drug development process with limited amount of patient data before the planning of Phase 3 clinical trials. Despite the challenge of small sample size, we pioneered the use of deep-learning derived digital pathology scores to identify responders based on the immunohistochemistry images of the target antigen expressed in tumor biopsy samples from a Phase 1 Non-small Cell Lung Cancer clinical trial. Based on repeated 10-fold cross validations, the deep-learning derived score on average achieved 4\% higher AUC of ROC curve and 6\% higher AUC of Precision-Recall curve comparing to the tumor proportion score (TPS) based clinical benchmark.  In a small independent testing set of patients, we also demonstrated that the deep-learning derived score achieved numerically at least 25\% higher responder rate in the enriched population than the TPS clinical benchmark.
\end{abstract}
\section{Introduction}
Traditionally, tumor proportion score (TPS) has been used as a histopathology image biomarker to identify responders. Although it was successfully used in multiple clinical development programs and contributed to many successful regulatory approved therapies, e.g. the recent breakthrough in immuno-oncology \citep{herbst2016pembrolizumab,reck2016pembrolizumab}, it was well recognized as an imperfect biomarker \citep{silva2018pd,piper2019can}. In \cite{silva2018pd}, the authors suggest the use of reactive vs constitutive PD-L1 staining patterns rather than TPS may lead to more accurate responder identification. Thus, to identify and quantify the expression patterns of targeted antigens exhibited in immunohistochemistry (IHC) images, we proposed to use deep learning based approaches to analyze IHC images to derive histopathology image scores for responder identification. However, there are three unique challenges: small sample size, high dimension and heterogeneity. First, since ideally the responder identification needs to be completed before Phase 3 clinical trials start, the  images available for deep learning is usually from much smaller clinical trials with less than a hundred patients. Second, the resolution of IHC images can be up to 60,000 by 60,000 because of 20 to 40 times magnification of tumor tissues. Last but not least, the whole slide images can be very heterogeneous due to the complicated IHC staining process and interpatient variability. To alleviate the above three challenges, we adopt the techniques of transfer learning, tiling, data augmentation and attention mechanism. 

 \paragraph{Related Work}
 Tabular data has been shown useful in predicting clinical trial results and identifying responders \citep{qi2019predicting}. However,  there has been very limited work in utilization of biomedical imaging data and deep learning for increasing probability of success of Phase 3 clinical trials. Most related work are in the field of healthcare with much larger sample size than that of a typical Phase 1 clinical trial. Patient outcomes observed out side of an clinical trial setting were predicted based on histopathology images. CNN (VGG-16) plus LSTM have been used in prediction of colorectal cancer patient 5-year survival based on H\&E slides from 420 patients and the method has achieved results better than human experts \citep{bychkov2018deep}. A modified VGG network has been used to predict survival risks of brain cancer patients based on H\&E slides from 769 patients \citep{mobadersany2018predicting}. Both of the above require a human expert to select a region of interest (ROI) for testing on an independent set of new patients and thus the generalizability may depend on how an ROI was selected and suffers from subjectivity.        

\section{Methodology}\label{sec:method}
To remove such subjectivity, the whole slide image needs to be fed into a pipeline with an automatic filtering mechanism so that the model can focus on regions of interest. To achieve that, several types of tumor specific staining approaches were experimented with additional cost to drug development program, since the drug considered has a mechanism of action targeting on an antigen expressed by tumor cells though not all tumor cells express such antigen. However, these staining were not 100\% accurate and often had false negatives and false positives \citep{togo2017sensitive}. Thus, a tumor recognition model is needed to directly identify tumor regions without further staining.  On the other hand, granular cell level annotation of tumor on the existing IHC images takes too much time and cost. The solution we proposed is to use weak labels, i.e. labels at tile level, to train a tumor recognition model. Figure \ref{fig:data_pre} illustrates the data preparation pipeline for training a tumor recognition model. For supervised learning with weak labels, we chose an attention based deep multiple instance learning (MIL) method \citep{ilse2018attention} because of its attention mechanism and strong performance achieved under small sample size setting. 

\begin{figure}[htbp]
  \floatconts
  {fig:data_pre}
  {\caption{Illustration of the data preparation pipeline for the tumor recognition modeling step and the responder identification modeling step. }}
  {\includegraphics[width=1\linewidth]{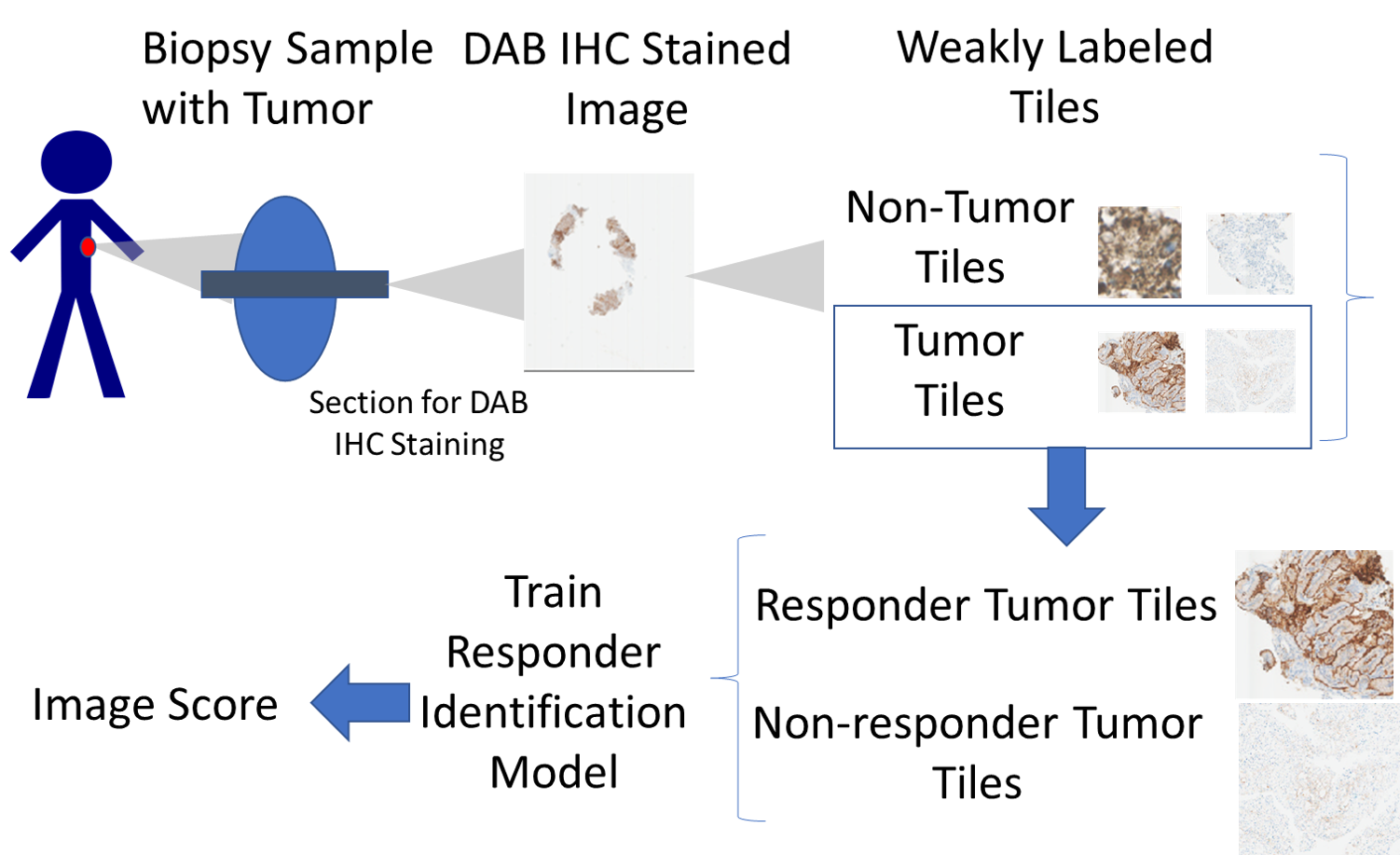}}

\end{figure}


We proposed a two step approach to derive image scores for the responder identification problem as shown in Figure \ref{fig:data_pre}. The first step is to train a tumor recognition model to recognize tumor tile vs non-tumor tile. For this step,  the  responder's and the non-responder's tiles were combined for training the model. The second step is to train a responder identification model to predict responding status of a patient based on tumor tiles only. For both steps, the transfer learning technique was used. The MobileNet \citep{howard2017mobilenets} architecture pretrained on ImageNet \citep{deng2009imagenet} was used because of the smaller size of the learnt features comparing to other pretrained networks and thus less likely to lead to overfitting issues than other much bigger network architectures for analyzing a small Phase 1 clinical trial.  

\section{Experiment}
To illustrate the performance of the proposed two-step MIL approach for responder identification in early phase clinical development, a cohort from a Phase 1 clinical trial in oncology was selected. The patients recruited into the cohort were from 26 clinical sites across five countries from three continents. For the reason of confidentiality, product information and drug target information are not disclosed in this manuscript. 

\textbf{Study Cohort}: The study cohort consists of 66 patients who were identified as intent-to-treat subject with an diaminobenzidine (DAB) stained IHC image taken at baseline and an evaluable best overall tumor response \citep{eisenhauer2009new} to an oncology treatment in a Phase 1 non-small cell lung cancer clinical trial. The best overall response can be dichotomized into responder and non-responder categories following the standard rule of RECIST 1.1 \citep{eisenhauer2009new} in tumor response evaluation, where complete response (CR) and partial response (PR) are grouped into the responder category while the rest grouped into the non-responder category. The modeling goal is to identify responders based on IHC images for increasing Phase 3 clinical trial's probability of success.      

\textbf{Data Preprocessing}: Due to the challenge of high resolution, each IHC image was first preprocessed to segment out the tissue region using the Otsu filter \citep{otsu1979threshold} and then each segmented tissue region was split into tiles with size $4096\times4096$, which is a size large enough to contain at least one tumor nest and often used in previous digital pathology literature \citep{bentaieb2018predicting,momeni2018deep}. The number of $4096\times4096$ tiles can range from one to hundreds for a patient depending on the size of the biopsy sample. 

\textbf{Data Augmentation}: To ensure equal representation of each patient in the responder identification step, a patient's tiles were augmented such that the total number of tiles per patient is the same as the patient with the largest number of tiles. 

\textbf{Tile Level Labeling}: After the above preprocessing, a tumor vs non-tumor label was assigned for each tile with the assistance from a pathologist while the responder vs non-responder label for each tumor tile was inherited from the patient level responder status.  

\textbf{Training Strategy}: Due to competitive landscape in oncology drug development, a Phase 3 trial was usually planned before completion of the Phase 1 clinical trial to gain advantages over competitors and bring novel therapies to patients faster. Thus, to inform patient selection for Phase 3 clinical trials, the modeling effort was started with the first 46 out of the 66 patients who had evaluable tumor response while the Phase 1 trial was still ongoing. Thus the first 46 patients, 10 of which are responders, were included into the training and validation process and the rest 20 patients, 4 of which are responders, were included into an independent testing set.
Since the cost of incorrectly classifying a responder as a non-responder is larger than that of incorrectly classifying a non-responder as a responder when there are few life saving treatments, the responders were weighted four times more than that of the non-responders during model training.  Although tumor tiles almost doubles the number of non-tumor tiles, to ensure tumor tiles can be correctly classified, tumor tiles and non-tumor tiles are equally weighted. 
To avoid the issue of the overfitting, the image data was randomly split  at the patient level into 80\% for training and 20\% for validation in the tumor recognition step and the responder identification step, respectively. For the responder identification step, because of a much smaller number of responders, a stratified splitting strategy was taken to make sure at least one responder was included into the validation set.  For the responder identification step, only the tiles with ground truth label of tumor tile were included into the training and validation sets.  In the testing stage, the tumor recognition model was used to filter out tiles predicted to be a non-tumor tile and only predicted tumor tiles were passed to the responder identification step.


An ablation study was conducted to compare a single step MIL approach vs the proposed two step MIL approach and examine the benefit of with or without data augmentation based on the average AUC of the ROC curves from a repeated 10-fold cross validation (CV) procedure, since CV is a standard procedure for model comparison. In standard CV, a performance metric, such as accuracy, was evaluated for each fold that serves as the validation set. However, due to the small sample size (about 4 subjects in each fold) we aggregate the predictions for each fold and evaluate a performance metric only after all the folds were completed. The modified CV procedure has been popular in the clinical trial subgroup identification settings with small sample size \citep{huang2017patient,chen2015prim}.  To reduce the randomness in the results, the 10-fold cross validation was repeated 10 times. 

\textbf{Clinical Benchmark}: The clinical benchmark is the tumor proportion score (TPS)  method for patience selection, which contributed to multiple successful development of novel oncology treatments \citep{reck2016pembrolizumab, migden2018pd,schmid2018atezolizumab}. TPS is calculated as the percentage of tumor cells that expressed a targeted antigen among all tumor cells. Patient with higher TPS is expected to have better response to the treatment. However, due to  complicated cell-cell interactions and heterogeneous expression patterns, not all patients with high TPS respond. Thus, it was well recognized that TPS is not a perfect biomarker for patient selection \citep{ilie2016assessment}.  

\section{Results} 
Different combinations of learning rate, number of neurons, activation functions were experimented and the a final set of hyperparameters in Table \ref{table:hyper} were chosen for fitting the final model.   

\subsection{Testing Results of the Final Model}
In the testing set, the MIL tumor identification model achieved over 95\%  precision-recall curve AUC and ROC AUC as shown in Table \ref{tab:tumor_results}. The responder identification model  achieved numerically at least 25\% higher precision (i.e. responder rate in the enriched population) than the TPS$\geq 50\%$ and TPS$\geq 75\%$ clinical benchmarks while maintaining the recall (i.e. percentage of responders correctly predicted among all true responders) same as or higher than the clinical benchmarks as shown in Table \ref{tab:responder_results}.

\begin{table}[htbp]
\floatconts
{tab:responder_results} 
  {\caption{Comparison between the two step MIL method and the TPS clinical benchmarks in an independent testing set. } }
  
  { \resizebox{\columnwidth}{!}{\begin{tabular}{| c |c| *{3}{p{2cm} |}}\hline
   Dataset & Method & Accuracy (95\%C.I.) &  Precision (95\%C.I.)
 &   Enriched Population Size \\ \hline
    Testing:  & TPS $\geq$ 50\% & 30\% (10\%,50\%) &  22\% (5\%,42\%)      &  n=18, \#PRs=4   \\
    n=20,  &   TPS $\geq$ 75\%& 50\% (30\%,70\%) & 25\% (0\%,50\%) &  n=12, \#PRs=3\\
           \#PRs=4   &  Two Step MIL  &  85\% (70\%,100\%) &   50\% (17\%,100\%)  & n=8, \#PRs=4\\
                                \hline
 \end{tabular}}}
  
\end{table}


To illustrate the effect of the attention mechanism, heat maps of the attention weighted images of a responder and a non-responder were shown in Figure \ref{fig:MIL_heatmap} together with their original images. Figure \ref{fig:MIL_heatmap} shows that not only tumor (dark color region) but also tumor micro-environment (light color region) are important for responder identification, which are consistent with recent medical literature on other oncology treatments \citep{mpekris2020combining}. The TPS method failed to differentiate the two patients because both of them have TPSTPS$\geq 90\%$. In contrasts to the TPS method, the two step MIL method was able to correctly differentiate the responder in (c) versus the non-responder in (a) based on the target expression patterns and also the cells in the tumor micro-environment.

\subsection{Ablation Study} 
\label{sec:abalation}
An ablation study was conducted to evaluate the impact of data augmentation and compare a single step MIL, which directly models all the tiles from an IHC slide treating both non-tumor tiles and tumor tiles equally, with the two step MIL method.  The results were tabulated in Table \ref{tab:ablation_results}. It suggests that data augmentation is critical to the success of the two step MIL method and it leads to significant improvement of ROC AUC and precision-recall curve AUC comparing to the TPS method. It also suggests that filtering out non-tumor tiles did not lead to significant improvement over single step MIL approach in this case study. However, as argued in the introduction section, since the treatment targets the antigen expressed on top of the tumor nest, it is very challenging to interpret results based on tiles without a tumor nest. Thus, two step MIL approach is still preferred over the single step approach, although it requires the initial effort of tile level annotation of tumor vs non-tumor tiles.  
 
\begin{table}[htbp]
\floatconts
  {tab:ablation_results}
  {\caption{Ablation study results for the responder identification step. Area under the curve (ROC) of ROC and precision-recall  curves (PRC) evaluated based on 10 repetitions of a modified 10-fold cross validation under different modeling strategies.} }
  {\resizebox{\columnwidth}{!}{\begin{tabular}{|c|r|r|r|r|}\hline
Metrics & \multicolumn{1}{|p{1.4cm}|}{\centering TPS \\ Baseline } & \multicolumn{1}{|p{2.5cm}|}{\centering Single \\ Step}  &  \multicolumn{1}{|p{2.5cm}|}{\centering  Two Step \\ Without \\ Augmentation}   &  \multicolumn{1}{|p{2.5cm}|}{\centering Two Step \\ with \\ Augmentation}  \\
   \hline
    \multicolumn{1}{|p{2cm}|}{\centering  AUC ROC \\ (95\% CI) }  & 76.1\%  & \multicolumn{1}{|p{2.5cm}|}{\centering  65.0\% \\ (62.2\%,67.8\%) }   &\multicolumn{1}{|p{2.5cm}|}{\centering  65.8\% \\ (62.6\%,69.0\%) } 
  & \multicolumn{1}{|p{2.5cm}|}{\centering   80.1\% \\ (78.4\%,81.8\%)  } \\           
    \multicolumn{1}{|p{2cm}|}{\centering  AUC PRC \\ (95\% CI) } &  58.5\%&  \multicolumn{1}{|p{2.5cm}|}{\centering  40.7\% \\ (34.7\%,46.6\%)  }  & \multicolumn{1}{|p{2.5cm}|}{\centering  41.7\% \\ (36.4\%,47.0\%) } 
  & \multicolumn{1}{|p{2.5cm}|}{\centering   64.7\% \\ (61.5\%,67.9\%) } \\         \hline                   
  \end{tabular}}}

\end{table}

\bibliography{biblio}
\appendix
\section*{Appendix A.}
\begin{figure}[htbp]
\floatconts
{fig:MIL_heatmap}
{\caption{ Visualization of the attention mechanism used in the two step MIL method.  (a) is a tumor tile of a non-responder and (b) is the attention weighted version of (a) with weights from the attention layer of the tumor recognition model.  (c) is a tumor tile from a responder and (d) is the attention weighted version of (c). In (b) and (d), the red color indicates a tile positively associated with a responder status while the blue color indicates the contrary. Darker tiles indicate tiles with smaller weights.}}
 {\includegraphics[width=0.8\linewidth]{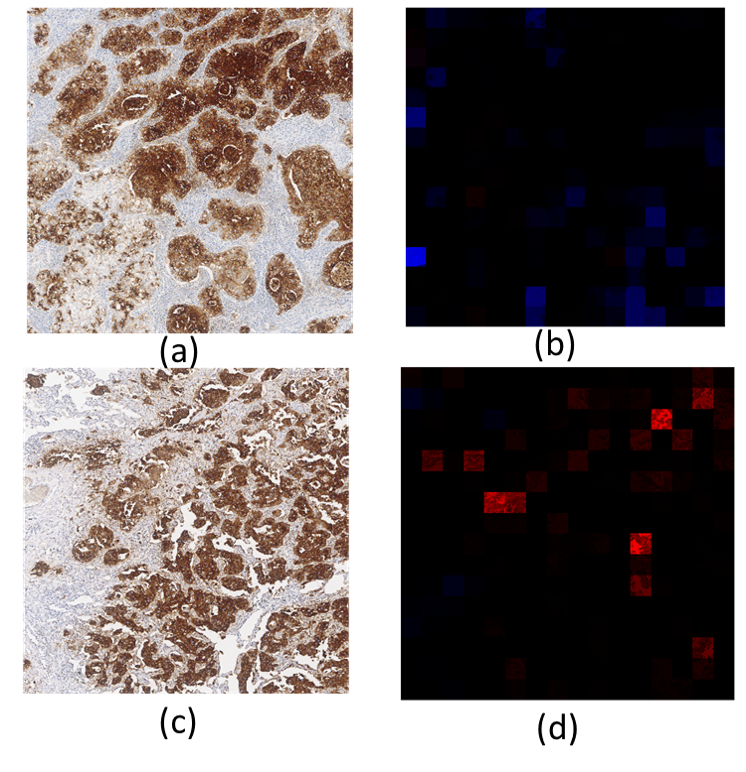}}
  
\end{figure}

\begin{table}[htbp]
\floatconts
    {tab:tumor_results}
   {\caption{Performance of the Deep Attention MIL method for the tumor recognition step in the testing set of 20 patients.} }
  {\begin{tabular}{|l|r|}\hline
   Metric  &     AUC(95\%C.I.)  \\ \hline
                                
    ROC     &   95.5\%
    (94.1\%,97.0\%) \\
    PRC &  99.0\%
    (98.7\%,99.4\%) \\
                                \hline
                                
  \end{tabular}}
\end{table}

The chosen hyperparameters of the final model are displayed in Table \ref{table:hyper}. 
\begin{table}[htbp]
\floatconts
{table:hyper}
{\caption{The chosen hyperparameters for the final model  }}
 {\resizebox{\columnwidth}{!}{\begin{tabular}{|c| c|} 
 \hline
 Hyperparameter & Value \\ [0.5ex] 
 \hline\hline
Learning rate & Cyclic Learning Rate 
(min: 0.00001, max: 0.001)\\
Optimizer & Adam\\
Number of Neurons in the FC Layers& 512\\
Number of FC Layers& 2\\
Number of Neurons in MIL Attention Layer & 128 \\
Activation Function & ReLU \\ [1ex] 
 \hline
 \end{tabular}}}
 
\end{table}

\end{document}